# Compounding Meta-atoms into Meta-molecules with Hybrid Artificial Intelligence Techniques

*Zhaocheng Liu[1,*], Dayu Zhu[1,*], Kyu-Tae Lee[1,*], Andrew S. Kim[1], Lakshmi Raju[1], Wenshan Cai[1,2]*

[1] School of Electrical and Computer Engineering, Georgia Institute of Technology, Atlanta, Georgia 30332

[2] School of Materials Science and Engineering, Georgia Institute of Technology, Atlanta, Georgia 30332

[*] These authors contributed equally to this work.

Correspondence to: wcai@gatech.edu

**Abstract**:
Molecules composed of atoms exhibit properties not inherent to their constituent atoms. Similarly, meta-molecules consisting of multiple meta-atoms possess emerging features that the meta-atoms themselves do not possess. Metasurfaces composed of meta-molecules with spatially variant building blocks, such as gradient metasurfaces, are drawing substantial attention due to their unconventional controllability of the amplitude, phase, and frequency of light. However, the intricate mechanisms and the large degrees of freedom of the multi-element systems impede an effective strategy for the design and optimization of meta-molecules. Here, we propose a hybrid artificial intelligence-based framework consolidating compositional pattern-producing networks and cooperative coevolution to resolve the inverse design of meta-molecules in metasurfaces. The framework breaks the design of the meta-molecules into separate designs of meta-atoms, and independently solves the smaller design tasks of the meta-atoms through deep learning and evolutionary algorithms. We leverage the proposed framework to design metallic meta-molecules for arbitrary manipulation of the polarization and wavefront of light. Moreover, the efficacy and reliability of the design strategy are confirmed through experimental validations. This framework reveals a promising candidate approach to expedite the design of large-scale metasurfaces in a labor-saving, systematic manner.

## 1. Introduction

Over the past two decades, the exploration of artificially structured optical media, such as plasmonics, metamaterials, and metasurfaces, has led to the discovery of exotic light-matter interactions and thus enabled vast applications of photonics ranging from highly integrated optical systems to advanced quantum photonic devices[1-6]. The anomalous capabilities of the metamaterials and metasurfaces are endowed by the building blocks, or "meta-atoms," with artificially engineered topology within the photonic materials. However, photonic media with identical meta-atoms may offer only inadequate controllability of light due to the limited flexibility rendered in periodic meta-structures of simple unit cells. To overcome these deficiencies, metasurfaces comprised of multiple meta-atoms, such as gradient and multi-layered metasurfaces, have been proposed and developed[7-9]. Relying on the collective effects of multiple meta-atoms, these metasurfaces present intriguing properties such as anomalous deflection[7, 10], arbitrary phase control, asymmetric polarization conversion[8, 11], etc., which brings about extensive applications for imaging, optical signal processing, emission control, and many more. Here in our following discussion, we refer to unit cells composed of various meta-atoms as meta-molecules, analogous to the hierarchical relationship between atoms and molecules in nature. In our definition of a meta-molecule, we assume every two adjacent meta-atoms are not strongly coupled, in which case the overall properties of the meta-molecule can be analytically predicted by the properties of its constituent meta-atoms. Such an assumption is valid in most metasurfaces that consist discrete, spatially variant building blocks.

Despite the extraordinary properties of metasurfaces made up of meta-molecules, designing multiple meta-atoms that collectively function as a device is a time-consuming task that requires labor-intensive trial-and-error simulations. The difficulty of the inverse design of such meta-molecules arises from the intricate mechanisms of multi-structured systems, the vast number of possible combinations of distinct meta-atoms, as well as the expensive three-dimensional full wave simulations required. Traditionally, a practical solution to such a design follows three steps: (1) specifying a class of geometry with a few parameters as candidate meta-atoms, (2) carrying out parametric sweeps on these parameters, and (3) enumerating possible combinations of meta-atoms to meet the design objective. However, the limitation of the geometry in the strategy largely restricts the variety of the shapes of meta-atoms, which usually does not lead to an optimal solution, even after extensive and expensive simulations.

Alongside the evolution of nanophotonics, various methods for expediting the design of photonic structures have been developed. Gradient-based adjoint methods, such as topology optimization, are a class of widely applied approaches for optimizing and designing nanophotonic devices[12-17]. To identify the optimized parameters of a device, the algorithm computes the gradient, or sensitivity, through the corresponding adjoint problem and updates the parameters along the deepest-gradient direction. In addition to adjoint methods, genetic algorithms and related variations also play important roles in the design of photonic structures[18-20]. The philosophy of the algorithms is to treat photonic structures as a population of individuals, and carry out bio-inspired operations like selection, reproduction, and mutation to the population in order to identify the optimized individual through evolution. Both gradient-based methods and genetic algorithms require immense amount of simulations. The slow convergence and local minimum problem of the two classes of approaches limit their applicability to the inverse design of nanostructures that require iterative and expensive full wave simulations. Recently, data-driven and deep learning-enabled approaches have been showing their powerful capacity in dealing with complex systems with extensively large degrees of freedom[21-23]. Once the dataset is constructed, the machine learning model trained based on the dataset can be employed for various designs in an expeditious way, multiple times. Applying deep learning approaches to the inverse design of photonics have effectively alleviated the problems of traditional design methods such as the slow convergence of optimization and the expensive cost of failure design[24-27]. In conjunction with traditional optimization techniques, it has been proved that deep learning can substantially mitigate problems such as the convergence to local minima and the curse of dimensionality in other optimization schema[23, 28-30].

Despite the development of techniques for the optimization of photonic structures, the inverse design of metasurfaces with meta-molecules, of which the degrees of freedom are astronomical, is still not resolved. Although the collective properties of meta-molecules can be predicted by individually simulating each meta-atom, the enormous number of possible combinations of candidate structures impedes efficient designing using state-of-the-art optimization techniques. In order to effectively discover and design functional meta-molecules consisting of spatially variant meta-atoms, in this paper, we propose a hybrid framework of a compositional pattern-producing network (CPPN)[31-32] and a cooperative coevolution (CC)[33-34] algorithm. The CPPN is an artificial generative network that can compose high-quality two-dimensional patterns in a pixel-by-pixel manner, while the CC is an evolutionary algorithm that is able to divide a large-scale optimization problem into subcomponents that can be tackled independently. In our hybrid framework, we utilize a CPPN to encode patterns representing the candidate meta-atoms into a low-dimensional latent space, and leverage the CC to identify a set of vectors in the latent space of the CPPN so that the corresponding meta-atoms can collectively function as the design objective. We have evaluated the performance of our framework by inversely designing metasurfaces made up of two classes of meta-molecules: (i) diatomic metasurfaces, which are comprised of diatomic meta-molecules for prescribed polarization control, and (ii) gradient metasurfaces, which consist of multiple meta-atoms for wavefront manipulation. The capacity of the proposed methodology is further verified via experimental characterizations of selected meta-molecules obtained from the inverse design framework. We note that the same framework can be applied to the inverse design of high-performance dielectric metasurfaces for other objectives with minimal alteration.

## 2. The hybrid framework

With the intention of designing meta-atoms and -molecules with arbitrary topologies, we first train a CPPN generator with a user-defined geometric dataset to produce nanostructure patterns with arbitrary shapes. Unlike convolutional neural networks (CNNs) that generate an entire image in one shot, CPPNs take the coordinates of the pixels in the image as an input one at a time, predict the corresponding pixel values, and, after iterating all the coordinates, assemble the predicted pixels as a pattern. As illustrated in **Fig. 1**, the inputs to the network are the coordinate tuple ($x$, $y$, $r$) and the bias vector $v$, where $r$ represents the distance of pixel ($x$, $y$) from the center of the image and $v$ is a parameter controlling the shapes of generated patterns. Here $v$ can be treated as the latent vector of the generated pattern since it is a low-dimensional vector representation of the high-dimensional composed image. Denoting the network as a function $f$ and the generated pattern as $s$, the network essentially performs the transform $s_i = f(x_i, y_i, r_i, v)$, where the subscript $i$ is the index of a pixel in the image. As the trained CPPN composes patterns in a pixel-by-pixel manner, it is able to generate patterns with arbitrarily complicated features including but not limited to corners, curves, and slits as desired by users. These fine features are crucial to strong resonance and coupling, and thus are indispensable to the discovery of novel photonic structures. We note that, given the geometric dataset containing patterns with proper geometric features, CPPNs are also able to compose patterns in a more randomized fashion for other inverse design problems. Implementation of a CPPN $f$ can be realized by any neural networks, and here we choose fully connected networks for simplicity. The architecture of the CPPN generator, the detailed training processes, and samples of generated patterns from the CPPN are presented in the Supporting Information.

In order to identify a meta-molecule with multiple meta-atoms while avoiding exponentially increased simulation time for the entire molecule, we introduce the cooperative coevolution (CC) into the framework. Specifically, we divide the design task of the whole meta-molecule into the design of independent meta-atoms, and iteratively optimize the latent vector of each meta-atom. As shown in the flowchart in **Fig. 1(b)**, the algorithm treats the population of latent vectors $v$ of meta-atoms as a species. In each iteration of the evolution, the CC algorithm picks one species for update while assuming all other species are optimized. The algorithm then decodes the latent vectors in the species through the CPPN, evaluates the fitness scores of the decoded meta-atoms, and optimizes the species by performing loops of bio-inspired operations including selection, reproduction, and mutation. These operation steps are also adopted in conventional evolution strategies. The CC algorithm performs the above steps and iterates in a round-robin fashion for all species (latent vectors of meta-atoms) until all their fitness scores reach the desired criteria. More detailed description of the algorithm, including the method for encoding nanostructures and the formulation of the fitness functions in different design tasks, are presented in the Supporting Information.

Our objective is to demonstrate the efficacy of our proposed algorithm for the inverse design of metasurfaces made up of meta-molecules for the manipulation of light. A general lattice schematic of such a metasurface is illustrated in **Fig. 2(a)**, where the meta-molecule, encircled by dashed lines, consists of multiple distinct meta-atoms indicated by blue, yellow, and orange circles. Figure 2(b) illustrates an example of the physical structure of a meta-molecule (diatomic meta-molecule) used in our design. The meta-molecule contains several gold meta-atoms patterned on a glass substrate. Each meta-atom has a thickness of $t = 40$ and is located in a square cell with a side length of $p = 320$ nm. The periodicity of the meta-molecule in the metasurface is $\Lambda = Np$, where $N$ is the total number of different meta-atoms. In order to accelerate the speed of our framework, we have trained a neural network simulator with 8,000 full wave simulations of meta-atoms of various shapes. The trained simulator is able to approximate the real and imaginary parts of the complex transmission and reflection coefficients over a prescribed spectral range, from 170 THz to 600 THz (i.e., 500 nm – 1.8 um in the wavelength domain) as used in the current study, with an accuracy above 97%.

## 3. Diatomic meta-molecules for polarization control

We first leverage our framework to design metasurfaces composed of "diatomic" nanostructures for polarization conversion. A diatomic meta-molecule by our definition is a meta-molecule consisting of two distinct meta-atoms. In such meta-molecules, the coupling between adjacent meta-atoms is sufficiently weak, in which case we can approximate the scattered far-field light from the meta-molecule as the superposition of the waves separately scattered by the two atoms. Based on this assumption, we inversely design a series of diatomic meta-molecules that are able to convert linearly polarized incident light into prescribed polarization states at the transmitted side. The representation of the light polarization and the method for computing fitness scores are detailed in the Supporting Information. **Figure 3** presents the designed meta-molecules that rotate the polarization angle of *x*-polarized incident light to 15°, 30°, 45°, and 60°, as well as the design that converts the incident light to a circularly polarized (CP) light, operated at a wavelength of 800 nm. Each design can be accomplished using our framework within 20 seconds on a machine with a single Quadro P5000 GPU. The plots of the converted polarization, computed by the network simulator and FEM full-wave simulation, are illustrated in Fig. 3(b) and 3(c) respectively. The intensities of the light are normalized to the maximum of each case to clearly display the polarization states; the plots of actual intensities are provided in the Supporting Information. The designed structures are able to accurately rotate or convert the polarization state, as demanded by the design objectives, in terms of both rotation angle and ellipticity.

We performed nanofabrication and optical characterizations of selected meta-molecules, as inversely designed using the hybrid artificial intelligence framework, to further validate the strategy we have developed. Specifically, here we present case studies for diatomic metasurfaces that can convert *x*-polarized (0°) incident light into linearly polarized (LP) light with polarization rotation angles of $\Psi = 30°$ and 45°. The SEM images of the two fabricated samples are shown in Fig. 3(d) and 3(g), while the measured polarization states (dashed lines) as compared to the simulated design objectives (solid lines) are provided in Fig. 3(e) and 3(h). The experimental results perfectly match the design objectives despite marginal disagreements incurred by error inherent to the fabrications process. To quantify the overall performance of the two diatomic meta-molecules, we simulate the transmitted polarization states denoted by the tuples $(\Psi, \chi)$, representing orientation and ellipticity of the polarization, and plot them on a Poincare sphere as illustrated in Fig. 3(i) and 3(f). The blue crosses represent the input polarization (horizontal), and the green/red ones denote simulated output polarization states after the metasurfaces. The designed metasurfaces with diatomic meta-molecules transform the horizontal polarization at the input to the output states with orientation angles of $\Psi = 31.3°$ and 43.7°, respectively, along with minimal ellipticities, which remarkably match the design objectives with marginal discrepancy. When the incident polarization direction $\Psi$ varies from 0° to 180°, which corresponds to the blue dots on the equator of the Poincare sphere, the polarization states are mapped to a great circle (green and red dots) slighted tilted from the equator. The detailed mapping relation between the input polarization states and the converted ones is provided in the Supporting Information.

## 4. Gradient metasurfaces

We further utilize our framework to inversely design metasurfaces with a gradient phase distribution. Such metasurfaces are conventional examples of metasurfaces for diverse functionalities such as the generalized Snell's law, beam steering, and meta-holography. Yet, no existing tools based on artificially intelligence have demonstrated their capacity to tackle this classical problem that requires the design of spatially variant, two-dimensional, arbitrarily shaped meta-atoms and their assemblies. The meta-molecules in our gradient metasurfaces for wavefront control and polarization manipulation are composed of eight meta-atoms. In such a meta-molecule, every two adjacent atoms should scatter light with an equal amplitude and a constant phase difference $\Delta\phi$. This requirement cannot be formulated into a single objective function to be minimized by traditional optimization techniques. To circumvent this problem, we design a fitness function of a meta-atom only associated with itself and its adjacent neighbor. For example, to optimize the $i^{th}$ meta-atom $s_i$ in the meta-molecule, we define its fitness function $F_i(s_i, s_{i-1})$ which is only locally associated with the $(i-1)^{th}$ meta-atom. As a greedy algorithm, the CC does not guarantee optimal solutions, but empirically, it converges to a solution with the desired amplitude requirement and phase distribution in our experiments. The full description of the fitness function is presented in the Supporting Information.

We apply our framework to the design of an eight-atom meta-molecule, which can convert $x$-polarized incident light to $y$-polarized counterpart, and deflect the cross-polarized light to a specific angle. We choose the central operating wavelength of the metasurface as 800 nm to facilitate further experimental verifications. Given such requirements, our framework is able to identify a set of meta-atoms satisfying the design objective within 10 minutes on a GPU machine. **Figure 4(a)** shows the unit cell of an identified metasurface, together with the SEM image of the corresponding fabricated sample. We conducted polarization analysis of the diffracted light on the transmitted side, as illustrated in Fig. 4(b), where the red dashed line indicates the measured polarization state. As compared to the incident polarization (blue dashed line), the converted polarization is perfectly orthogonal to the incident one as demanded by the design objective. Figure 4(c) presents the FEM simulated distribution of the electric field $E_y$ emerging from the metasurface, which further illustrates the accurate phase gradient and amplitude distribution induced by the eight meta-atoms.

Another example we present here is the inverse design and experimental verification of a gradient metasurface for circularly polarized (CP) light. The envisioned metasurface should be able to convert left circularly polarized (LCP) light to its cross-polarization (RCP) and bring about an additional phase gradient to the converted portion of light for beam steering. At the design wavelength of 800 nm, one of the identified solutions and its fabricated sample are shown in Fig. 4(d). We characterize the polarization states of the incident (blue) and converted (red) light in the polar graph as in Fig. 4(e), where the dash-dotted and dashed lines indicate the measured intensity with and without a quarter waveplate, respectively. The orthogonality of the polarizations measured with a quarter waveplate unambiguously proves the sample flips the circular polarization from LCP to RCP. We also provide the simulated converted electric field $E_{RCP}$ under the LCP incidence, confirming the correct polarization, phase gradient, and amplitude distribution of the design. Unlike traditional circular gradient metasurfaces, where only geometric phase contributes to the phase distribution, the phase gradient of our design is jointly induced by the geometric phase $\varphi_{geo}$ and the material-induced phase delay $\varphi_{mat}$. When the incidence is LCP, $\varphi_{geo}$ and $\varphi_{mat}$ constructively contribute to the desired phase gradient; however, under the RCP incidence, $\varphi_{geo}$ and $\varphi_{mat}$ destructively contribute to the phase profile, causing asymmetrical behavior of the metasurface. To illustrate this fact, in Fig. 4(g) we provide the phase delay induced by the eight meta-atoms under circularly polarized incidence. Unlike the case of LCP incidence (blue markers, which is the polarization this metasurface is designed to serve), where the phase delay parallels the linear distribution of a conventional circular gradient metasurface, the phase shift with the RCP incidence (orange markers) deviates from the constant phase gradient, causing multiple diffractions. This property is also reflected in the calculated far-field angular distributions presented in Fig. 4(h) and 4(i). Since our framework does not consider the particular physical mechanism during the inverse design process, nor does it require predefined constraints or human intervention, it tends to discover metaphotonic devices and novel photonic phenomena with complex light-matter interactions. More examples of inversely designed gradient metasurfaces for various operating wavelengths and polarization states are provided in the Supporting Information.

## 5. Discussion and Conclusion

We have proposed a framework consolidating a CPPN generator and a cooperative coevolution algorithm to resolve the inverse design of meta-molecules comprised of spatially variant meta-atoms. The CPPN generator creates high-quality nanostructure patterns with complicated features, while the CC algorithm expedites the identification of the multi-element metasurfaces by dividing the overall design problem into individual subtasks. With the acceleration of a neural network simulator, our framework is able to discover novel metasurfaces with prescribed design objectives, as exemplified in the inverse design and experimental verification of gradient metasurfaces for polarization and wavefront control. Since the proposed framework relies on stochastic evolutionary algorithms, it tends to converge to a solution near the global minimum under several, if not one, fast runs with random initializations. The cost of failure design is also dramatically reduced due to the inexpensive simulation enabled by the neural network simulator. As demonstrated in our paper, a well-trained neural network simulator is reliable enough for the practical application of the framework, but we would encourage to replace the simulator with a physical simulation approach at the last few optimization iterations. This will eliminate occasional flawed solutions incurred by the inaccurate approximation of the neural network simulator.

In addition to the case studies we have demonstrated here, the proposed strategy can be applied to the design of a wide range of metasurfaces and photonic devices whose overall performance can be predicted by the properties of their subcomponents such as multi-layered metasurfaces and achromatic metalenses. Thanks to the fast inference speed of neural networks, our algorithm is also capable of exploring unconventional optical phenomena induced by proper combination of multiple meta-structures with little cost of failed discoveries. In terms of the framework itself, as the ES optimization does not require gradients from the fitness function, the formulated design objectives can be arbitrary with little constraints. This unique feature enables the framework to be a powerful tool for general inverse design tasks aiming at on-demand, user-defined design objectives in terms of scattering, dispersion, chirality, absorption, nonlinear responses, and nearfield enhancement. Furthermore, since the proposed framework is an implicit method without consideration of the actual dynamics of the systems, the strategy developed here presents a broad and versatile potential for the exploration of novel phenomena and prototypes in generic design problems in other disciplines, such as materials science, condensed matter physics, and chemical syntheses.

**Figures and captions**

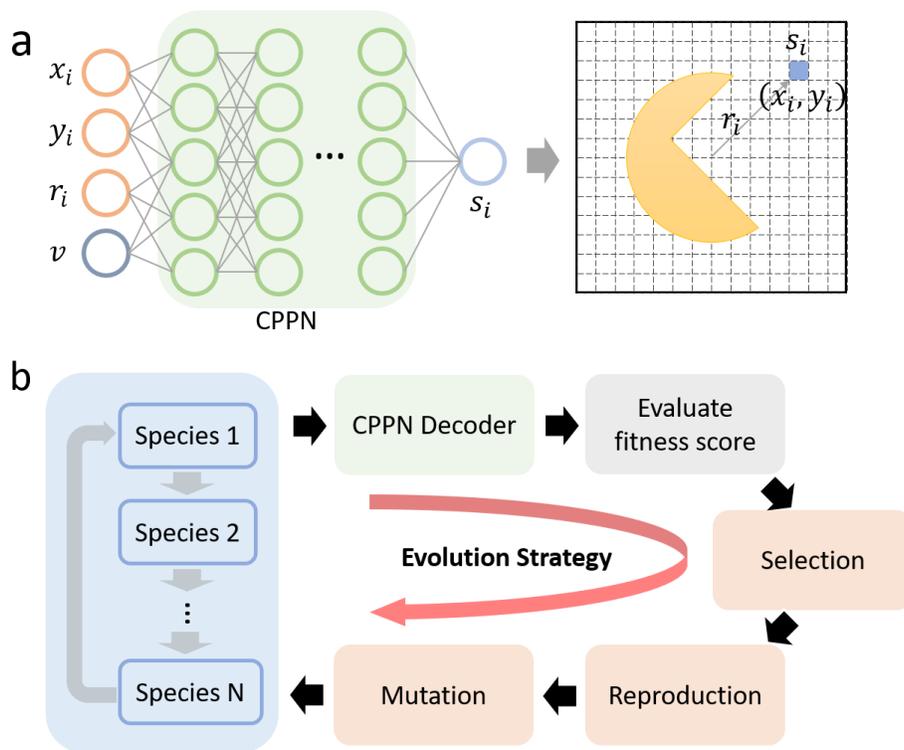

**Fig. 1 | Description of modules of the hybrid framework.** (a) The schematic of a CPPN. The network takes coordinates of a pixelated image $(x_i, y_i, r_i)$ and a bias vector $v$ as the input, and predicts the value $s_i$ of each pixel through the fully connected neural network. The predicted pixels are assembled to compose the pattern that represents a meta-atom. In this schematic, we use a single node to denote the bias vector, which, in practice, can be in arbitrary dimensions. (b) The flowchart of the cooperative coevolution algorithm. In each iteration, a species is selected and decoded into meta-atoms. The algorithm then evaluates the scores of all individuals in this species, and performs loops of bio-inspired operations such as selection, reproduction, and mutation. The algorithm iterates in a round-robin fashion until the fitness scores of all species meet the preset criteria.

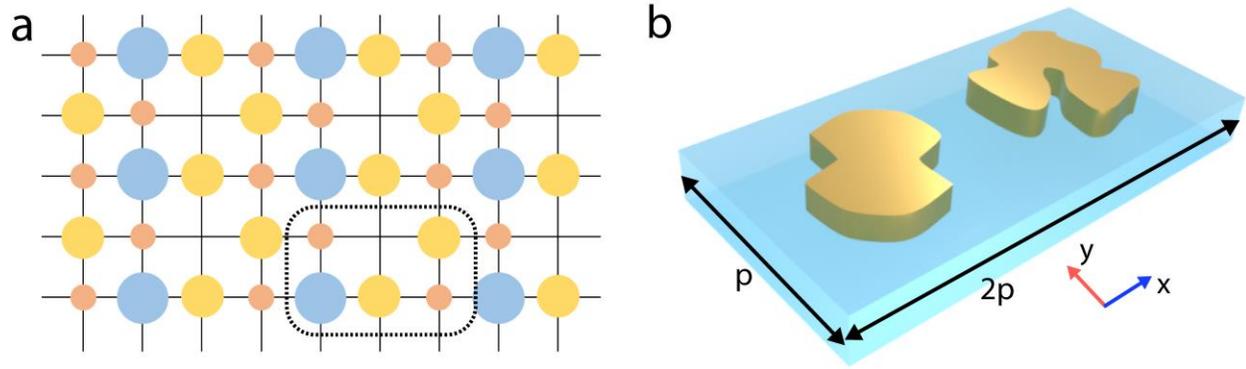

**Fig. 2 | Illustration of a meta-molecule in the design problem.** (a) The schematic of a two-dimensional lattice of a metasurface, which consists of arrays of meta-molecules. Each meta-molecule, as encircled by the dashed line, contains several distinct meta-atoms. (b) An example of a diatomic meta-molecule comprised of two meta-atoms. In the subsequent studies, the meta-atoms are gold particles with various topologies situated on glass substrate. Common parameters in the design problem include the thickness of the meta-atom, $t = 40$ nm, and the side length of the square lattice grid, $p = 320$ nm.

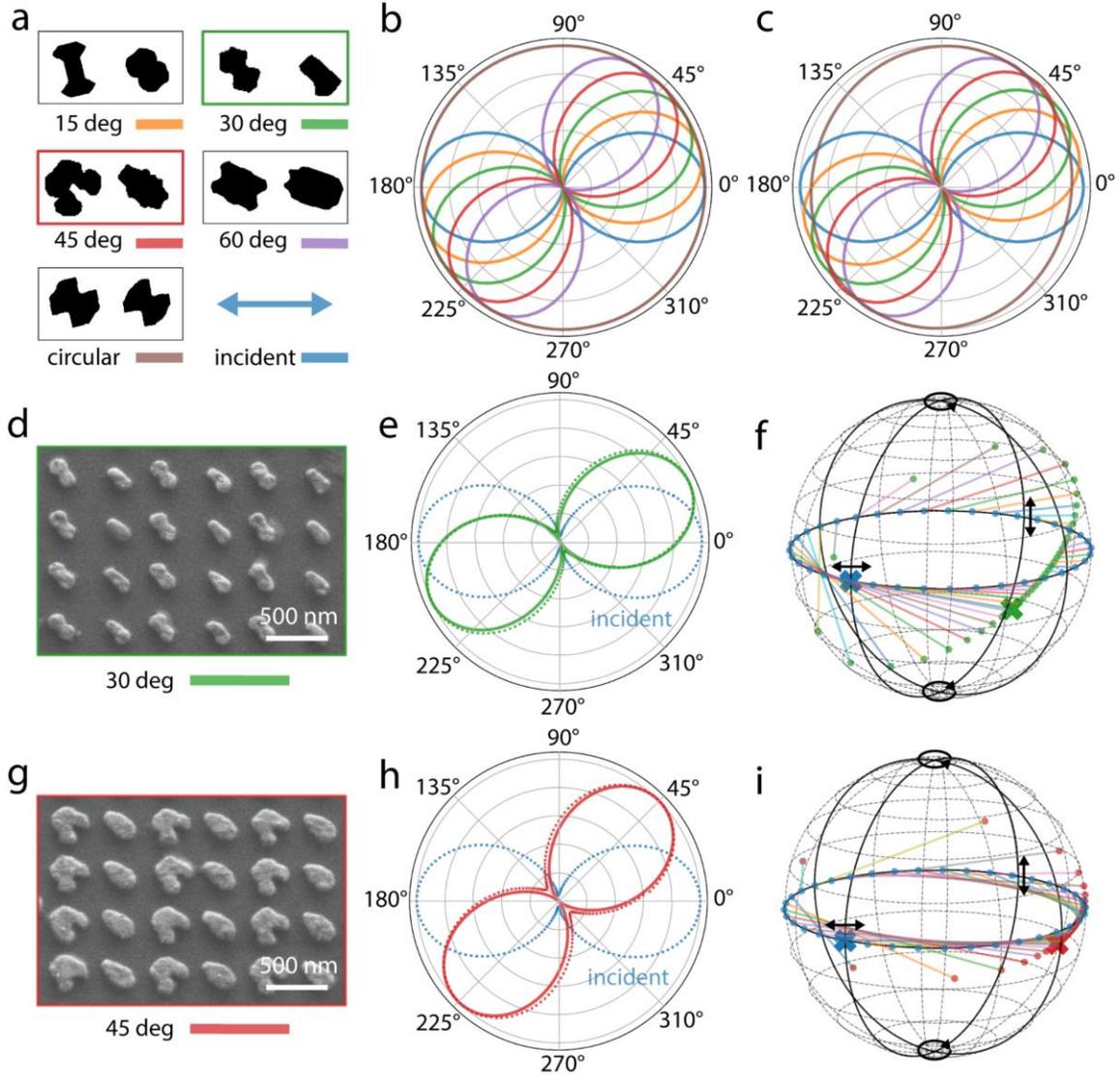

**Fig. 3 | Inversely designed diatomic meta-molecules for polarization manipulation.** (a) Unit cells of designed diatomic meta-molecules. The design objectives are prescribed polarization rotation/conversion for the transmitted light as indicated underneath each meta-molecule. (b – c) Simulated polarization states of transmitted light from the meta-molecules in (a) with a neural network simulator and FEM full-wave simulation, respectively. (d) and (g) SEM images of the fabricated metasurface polarization rotators with prescribed rotation angles of 30° and 45°, respectively. (e) and (h) Measured polarization states (dashed green/red lines) at the output of the metasurfaces, along with the state of the incident polarization (blue dashed lines at 0°). The solids lines represent the design objectives. (f) and (i) Incident and converted polarization states represented on Poincare spheres. The blue cross indicates the incident horizontal polarization, while the green and red crosses represent the states after the polarization conversion. The transmitted polarization states (green and red dots) at various incident polarization angles (blue dots) are also plotted on the spheres.

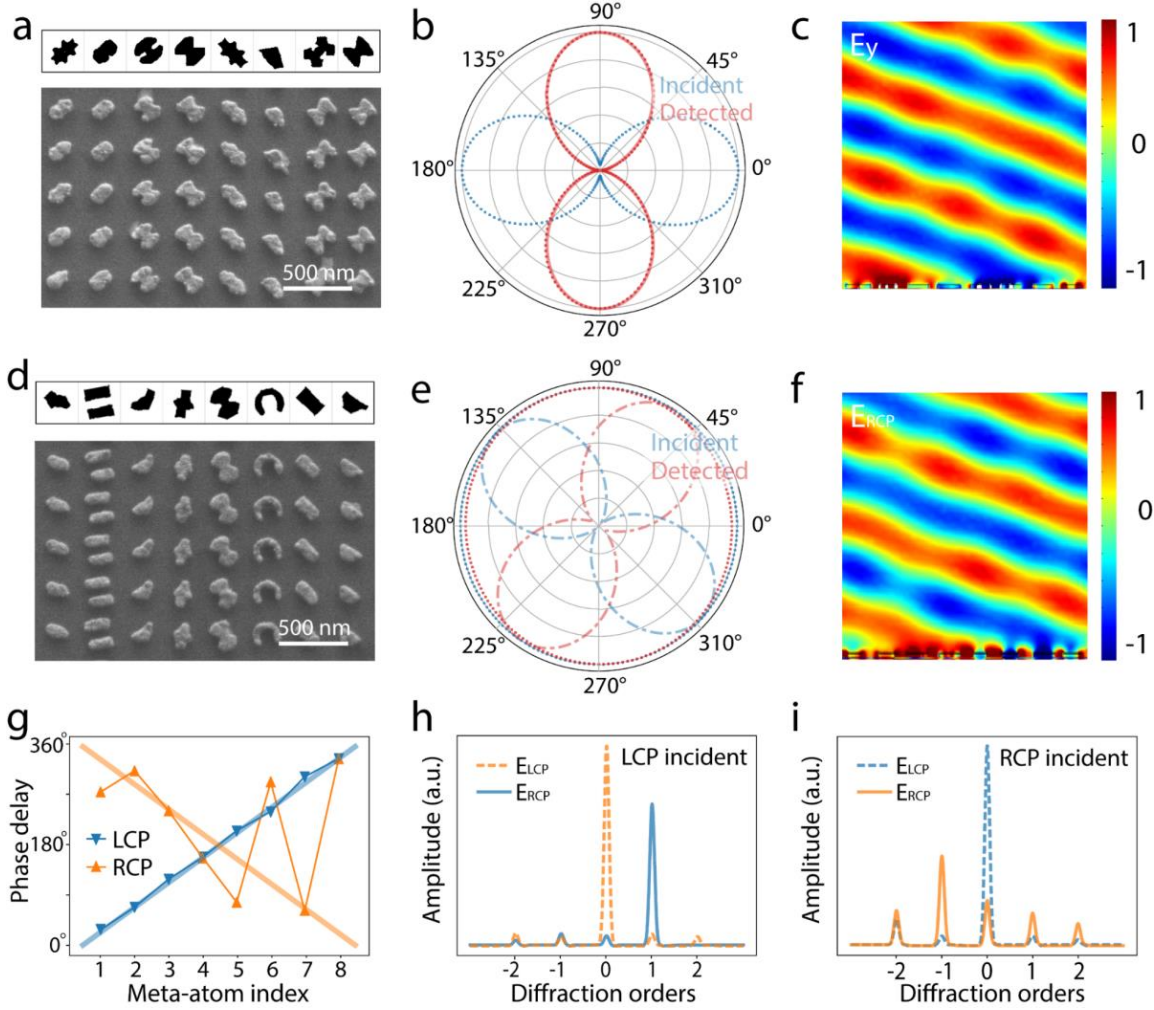

**Fig. 4 | Inversely designed meta-molecules for gradient metasurfaces.** (a) Unit cells and SEM image of a gradient metasurface for polarization conversion and beam steering with linearly polarized incidence. (b) Measured polarization states of the incidence (blue dashed line) and the diffracted beam (red dashed line). The solid red line (which essentially overlaps with the experimental data) indicates the design objective. (c) Simulated electric field $E_y$ distribution in the proximity of the metasurface shown in (a) under *x*-polarized incident light. (d) Designed gradient metasurface for polarization conversion and beam steering with LCP incidence. (e) Measured polarization states (dashed lines) of the incident and diffracted lights. The dash-dotted lines represent measured data after a quarter waveplate. (f) Simulated electric field $E_{RCP}$ distribution under the LCP incidence. (g) Relative phase delay induced by the eight meta-atoms shown in (d) for LCP and RCP incident waves, respectively. The solid lines represent the phase distribution of conventional geometric-phase-based circular metasurfaces. (h – i) Simulated diffraction behavior of the metasurface under LCP and RCP incident lights, respectively, where circularly polarizations of opposite handedness are steered to different directions. Note that the LCP incidence is the designed operating mode for this metasurface.

**Reference**
1. Barik, S.; Karasahin, A.; Flower, C.; Cai, T.; Miyake, H.; DeGottardi, W.; Hafezi, M.; Waks, E., A topological quantum optics interface. *Science* **2018,** *359* (6376), 666-668.
2. Mekis, A.; Chen, J.; Kurland, I.; Fan, S.; Villeneuve, P. R.; Joannopoulos, J., High transmission through sharp bends in photonic crystal waveguides. *Physical Review Letters* **1996,** *77* (18), 3787.
3. Cai, W.; Shalaev, V. M., *Optical metamaterials*. Springer: 2010; Vol. 10.
4. Fang, Y.; Sun, M., Nanoplasmonic waveguides: towards applications in integrated nanophotonic circuits. *Light: Science & Applications* **2015,** *4* (6), e294.
5. Li, G.; Zhang, S.; Zentgraf, T., Nonlinear photonic metasurfaces. *Nature Reviews Materials* **2017,** *2* (5), 17010.
6. Ni, X.; Wong, Z. J.; Mrejen, M.; Wang, Y.; Zhang, X., An ultrathin invisibility skin cloak for visible light. *Science* **2015,** *349* (6254), 1310-1314.
7. Yu, N.; Capasso, F., Flat optics with designer metasurfaces. *Nature Materials* **2014,** *13* (2), 139.
8. Pfeiffer, C.; Zhang, C.; Ray, V.; Guo, L. J.; Grbic, A., High performance bianisotropic metasurfaces: asymmetric transmission of light. *Physical Review Letters* **2014,** *113* (2), 023902.
9. Grady, N. K.; Heyes, J. E.; Chowdhury, D. R.; Zeng, Y.; Reiten, M. T.; Azad, A. K.; Taylor, A. J.; Dalvit, D. A.; Chen, H.-T., Terahertz metamaterials for linear polarization conversion and anomalous refraction. *Science* **2013,** *340* (6138), 1304-1307.
10. Wang, S.; Wu, P. C.; Su, V.-C.; Lai, Y.-C.; Chen, M.-K.; Kuo, H. Y.; Chen, B. H.; Chen, Y. H.; Huang, T.-T.; Wang, J.-H., A broadband achromatic metalens in the visible. *Nature Nanotechnology* **2018,** *13* (3), 227.
11. Schwanecke, A.; Fedotov, V.; Khardikov, V.; Prosvirnin, S.; Chen, Y.; Zheludev, N., Nanostructured metal film with asymmetric optical transmission. *Nano Letters* **2008,** *8* (9), 2940-2943.
12. Molesky, S.; Lin, Z.; Piggott, A. Y.; Jin, W.; Vucković, J.; Rodriguez, A. W., Inverse design in nanophotonics. *Nature Photonics* **2018,** *12* (11), 659.
13. Jensen, J. S.; Sigmund, O., Topology optimization for nano-photonics. *Laser & Photonics Reviews* **2011,** *5* (2), 308-321.
14. Borel, P. I.; Harpøth, A.; Frandsen, L. H.; Kristensen, M.; Shi, P.; Jensen, J. S.; Sigmund, O., Topology optimization and fabrication of photonic crystal structures. *Optics Express* **2004,** *12* (9), 1996-2001.
15. Sell, D.; Yang, J.; Doshay, S.; Yang, R.; Fan, J. A., Large-angle, multifunctional metagratings based on freeform multimode geometries. *Nano letters* **2017,** *17* (6), 3752-3757.
16. Lin, Z.; Groever, B.; Capasso, F.; Rodriguez, A. W.; Lončar, M., Topology-optimized multilayered metaoptics. *Physical Review Applied* **2018,** *9* (4), 044030.
17. Piggott, A. Y.; Lu, J.; Lagoudakis, K. G.; Petykiewicz, J.; Babinec, T. M.; Vučković, J., Inverse design and demonstration of a compact and broadband on-chip wavelength demultiplexer. *Nature Photonics* **2015,** *9* (6), 374.

18. Shen, L.; Ye, Z.; He, S., Design of two-dimensional photonic crystals with large absolute band gaps using a genetic algorithm. *Physical Review B* **2003,** *68* (3), 035109.
19. Wang, C.; Yu, S.; Chen, W.; Sun, C., Highly efficient light-trapping structure design inspired by natural evolution. *Scientific Reports* **2013,** *3*, 1025.
20. Preble, S.; Lipson, M.; Lipson, H., Two-dimensional photonic crystals designed by evolutionary algorithms. *Applied Physics Letters* **2005,** *86* (6), 061111.
21. Segler, M. H.; Preuss, M.; Waller, M. P., Planning chemical syntheses with deep neural networks and symbolic AI. *Nature* **2018,** *555* (7698), 604.
22. Carrasquilla, J.; Melko, R. G., Machine learning phases of matter. *Nature Physics* **2017,** *13* (5), 431.
23. Sanchez-Lengeling, B.; Aspuru-Guzik, A., Inverse molecular design using machine learning: Generative models for matter engineering. *Science* **2018,** *361* (6400), 360-365.
24. Liu, D.; Tan, Y.; Khoram, E.; Yu, Z., Training deep neural networks for the inverse design of nanophotonic structures. *ACS Photonics* **2018,** *5* (4), 1365-1369.
25. Ma, W.; Cheng, F.; Liu, Y., Deep-learning-enabled on-demand design of chiral metamaterials. *ACS Nano* **2018,** *12* (6), 6326-6334.
26. Peurifoy, J.; Shen, Y.; Jing, L.; Yang, Y.; Cano-Renteria, F.; DeLacy, B. G.; Joannopoulos, J. D.; Tegmark, M.; Soljačić, M., Nanophotonic particle simulation and inverse design using artificial neural networks. *Science Advances* **2018,** *4* (6), eaar4206.
27. Liu, Z.; Zhu, D.; Rodrigues, S. P.; Lee, K.-T.; Cai, W., Generative model for the inverse design of metasurfaces. *Nano Letters* **2018,** *18* (10), 6570-6576.
28. Liu, Z.; Raju, L.; Zhu, D.; Cai, W., A hybrid strategy for the discovery and design of photonic nanostructures. *arXiv preprint arXiv:1902.02293* **2019**.
29. Kudyshev, Z. A.; Kildishev, A. V.; Shalaev, V. M.; Boltasseva, A. In *Machine-learning-assisted topology optimization for highly efficient thermal emitter design*, CLEO: QELS_Fundamental Science, Optical Society of America: 2019; p FTh3C. 2.
30. Jiang, J.; Sell, D.; Hoyer, S.; Hickey, J.; Yang, J.; Fan, J. A., Freeform diffractive metagrating design based on generative adversarial networks. *arXiv preprint arXiv:1811.12436* **2018**.
31. Stanley, K. O., Compositional pattern producing networks: A novel abstraction of development. *Genetic Programming and Evolvable Machines* **2007,** *8* (2), 131-162.
32. Stanley, K. O.; D'Ambrosio, D. B.; Gauci, J., A hypercube-based encoding for evolving large-scale neural networks. *Artificial Life* **2009,** *15* (2), 185-212.
33. Yang, Z.; Tang, K.; Yao, X. In *Multilevel cooperative coevolution for large scale optimization*, 2008 IEEE Congress on Evolutionary Computation (IEEE World Congress on Computational Intelligence), IEEE: 2008; pp 1663-1670.
34. Potter, M. A.; Jong, K. A. D., Cooperative coevolution: An architecture for evolving coadapted subcomponents. *Evolutionary Computation* **2000,** *8* (1), 1-29.